\documentclass[aps,prx,a4paper,twocolumn,superscriptaddress,citeautoscript]{revtex4-1}
\usepackage{soul}
\usepackage[normalem]{ulem}
\usepackage{hyperref}
\usepackage{chngcntr}
\usepackage{color}
\usepackage[english]{babel}
\usepackage{lmodern}
\usepackage[T1]{fontenc}
\usepackage{amsmath}
\usepackage{units}
\usepackage[pdftex]{graphicx}
\usepackage{grffile}
\usepackage{color}
\usepackage[bottom]{footmisc}

\begin{document}
\title{Raman spectroscopic evidence for multiferroicity in rare earth nickelate single crystals}
\author{\href{https://orcid.org/0000-0001-6527-4035}{I.~Ardizzone}}
\affiliation{Department of Quantum Matter Physics, University of Geneva, 24  Quai Ernest-Ansermet, 1211 Geneva 4, Switzerland}
\author{\href{https://orcid.org/0000-0002-7590-2987}{J.~Teyssier}}
\affiliation{Department of Quantum Matter Physics, University of Geneva, 24  Quai Ernest-Ansermet, 1211 Geneva 4, Switzerland} 
\author{\href{https://orcid.org/0000-0002-1288-9236}{I.~Crassee}}
\affiliation{Department of Quantum Matter Physics, University of Geneva, 24  Quai Ernest-Ansermet, 1211 Geneva 4, Switzerland}
\author{\href{https://orcid.org/0000-0001-9574-6435}{A. B.~Kuzmenko}}
\affiliation{Department of Quantum Matter Physics, University of Geneva, 24  Quai Ernest-Ansermet, 1211 Geneva 4, Switzerland}
\author{\href{https://orcid.org/0000-0002-0421-0625}{D.~G.~Mazzone}}
\affiliation{Paul Scherrer Institut, PSI, Forschungsstrasse 111, 5232 Villigen, Switzerland}
\author{\href{https://orcid.org/0000-0003-4460-7106}{D.~J.~Gawryluk}}
\affiliation{Paul Scherrer Institut, PSI, Forschungsstrasse 111, 5232 Villigen, Switzerland}
\author{\href{https://orcid.org/0000-0003-1588-2870}{M.~Medarde}}
\affiliation{Paul Scherrer Institut, PSI, Forschungsstrasse 111, 5232 Villigen, Switzerland}
\author{\href{https://orcid.org/0000-0001-5266-9847}{D.~van~der~Marel}}\email{dirk.vandermarel@unige.ch}
\affiliation{Department of Quantum Matter Physics, University of Geneva, 24  Quai Ernest-Ansermet, 1211 Geneva 4, Switzerland}
\date{\today}
\begin{abstract}
The rare earth nickelates RNiO$_3$ are metallic at high temperatures and insulating and magnetically ordered at low temperatures. The low temperature phase has been predicted to be type II multiferroic, {\em i.e.} ferroelectric and magnetic order are coupled and occur simultaneously. Confirmation of those ideas has been inhibited by the absence of experimental data on single crystals. Here we report on Raman spectroscopic data of RNiO$_3$ single crystals (R = Y, Er, Ho, Dy, Sm, Nd) for temperatures between 10~K and 1000~K. Entering the magnetically ordered phase we observe the appearance of a large number of additional vibrational modes, implying a breaking of inversion symmetry expected for multiferroic order. 
\end{abstract}
\maketitle
\section{Introduction}
During recent years materials with the perovskite structure have become subject of intensive research in the field of photovoltaic cells~\cite{kalaiselvi2018}, battery engineering~\cite{xu2019} and cancer therapy~\cite{staedler2014}. 
In particular the nickelates with the general formula RNiO$_3$ (with R a trivalent rare earth) are good candidates for these applications by virtue of the interdependence in these compounds of the structural ~\cite{alonso2001,medarde2009,johnston2014,green2016,gawryluk2019,rana2018}, electronic~\cite{king2014,dhaka2015,hampel2019,georgescu2019,liu2019} and magnetic~\cite{garcia1994,rodriguez1998,fernandez2001,munoz2009,buitrago2015,ruppen2017,hampel2017,fomichev2020} order parameters. 
The most characteristic property of RNiO$_3$ is the transition from a high temperature metallic phase to a low temperature insulating phase. 
The temperature of this transition, $T_{MI}$, depends strongly on the radius of the $R^{3+}$ ion: for R from Lu to Pr the cell parameter increases almost 10 \%, while $T_{MI}$ reduces from 600~K (Lu) to 135~K (Pr). 
In the high temperature metallic phase the unit cell contains four formula units, where the four nickel atoms occupy equivalent lattice positions corresponding to identical electronic configurations. 
Upon entering the insulating phase a differentiation occurs in the local electronic configuration of the nickel atoms, leading to a modified unit cell containing two pairs of non-equivalent nickel atoms characterized by short and long bonds with the surrounding oxygen atoms~\cite{mizokawa2000,mazin2007,park2012,subedi2015,ruppen2015}. 
This phenomenology is now widely agreed upon and has been labeled in the literature as, among other things, {\textquotedblleft}bond disproportionation{\textquotedblright} and  {\textquotedblleft}breathing distortion{\textquotedblright}, the latter of which we employ in this paper. 
This is accompanied by a slight opening of the angle between $a$ and $c$ axis away from 90 degrees of the order of $0.1$ degree~\cite{gawryluk2020} corresponding to the monoclinic structure ($P2_1/n$), whereas the metallic phase is orthorhombic ($Pbmn$).
For the larger Nd and Pr ions the insulating state is magnetically ordered up to $T_{MI}$~\cite{palina2017,post2018,bisht2017,scagnoli2006}.  For RNiO$_3$ with the smaller rare earth atoms (R = Lu, Y, Er, Ho, Dy, Sm) the Ni atoms order magnetically at a temperature $T_N$ (the N\'eel temperature) well below $T_{MI}$ and the changes in the lattice parameters at $T_{MI}$ are less pronounced than with Pr and Nd. 
\begin{figure}[!b]
\begin{center}
\includegraphics[width=1.0\columnwidth]{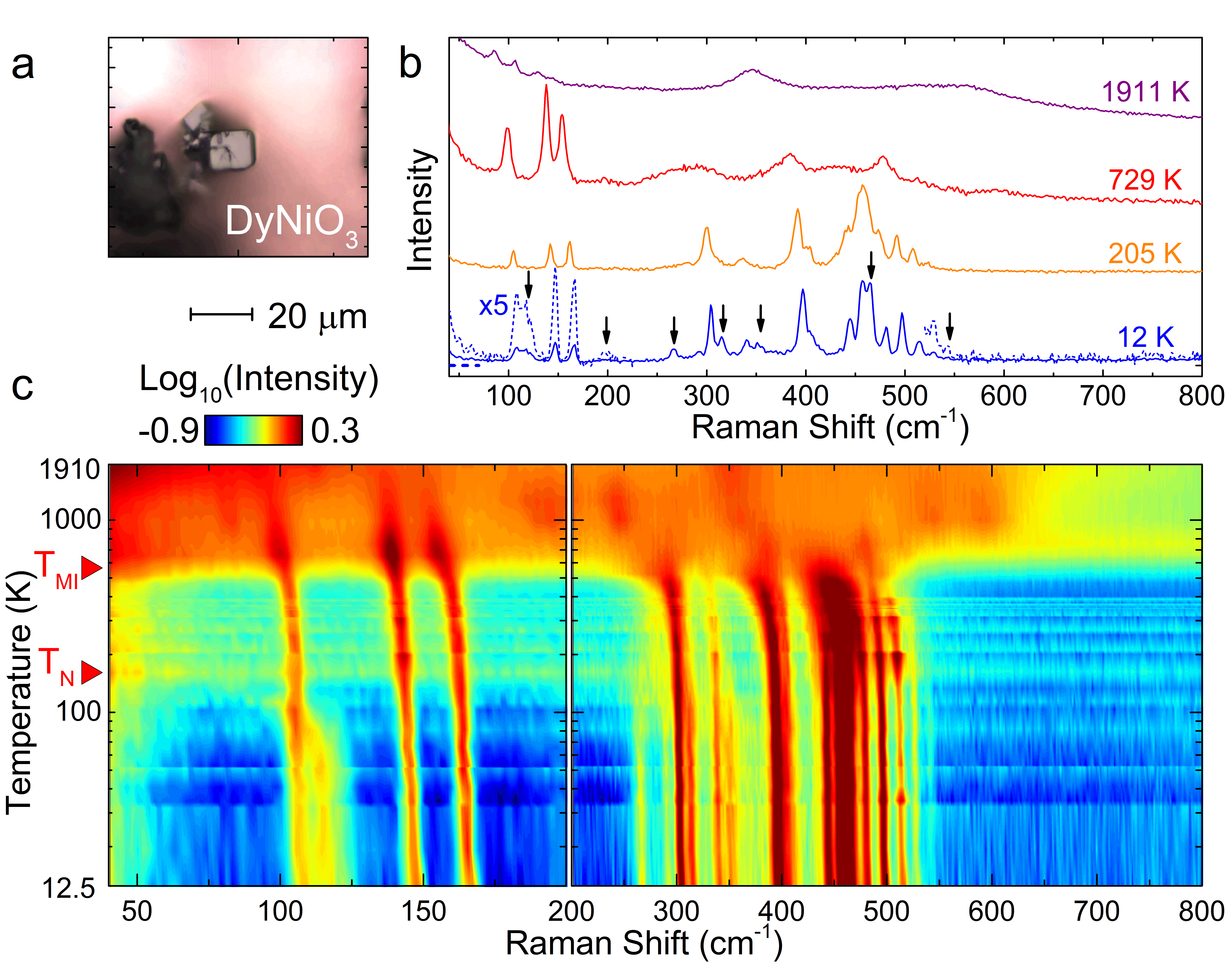}
\caption{\label{fig:1}
{\bf a}, Image of a DyNiO$_3$ single crystal inside the cryostat. 
{\bf b}, typical Raman spectra recorded on DyNiO$_3$ at temperatures in the antiferromagnetic insulating phase (blue curve, dashed line is a factor 5 magnification of the blue curve), paramagnetic insulating phase (orange curve) and metallic paramagnetic phase (red curve). 
The purple curve is close to the melting point of the material.
The curves have been given vertical offsets to avoid clutter. 
Raman modes appearing in the antiferromagnetic phase are marked by arrows. 
{\bf c}, colormap of the Raman response in the full temperature range from 12.5~K to 1911~K. 
Metal-insulator and magnetic transitions are labeled as red symbols. }
\end{center}
\end{figure}
\begin{figure*}[!t]
\begin{center}
\includegraphics[width=2\columnwidth]{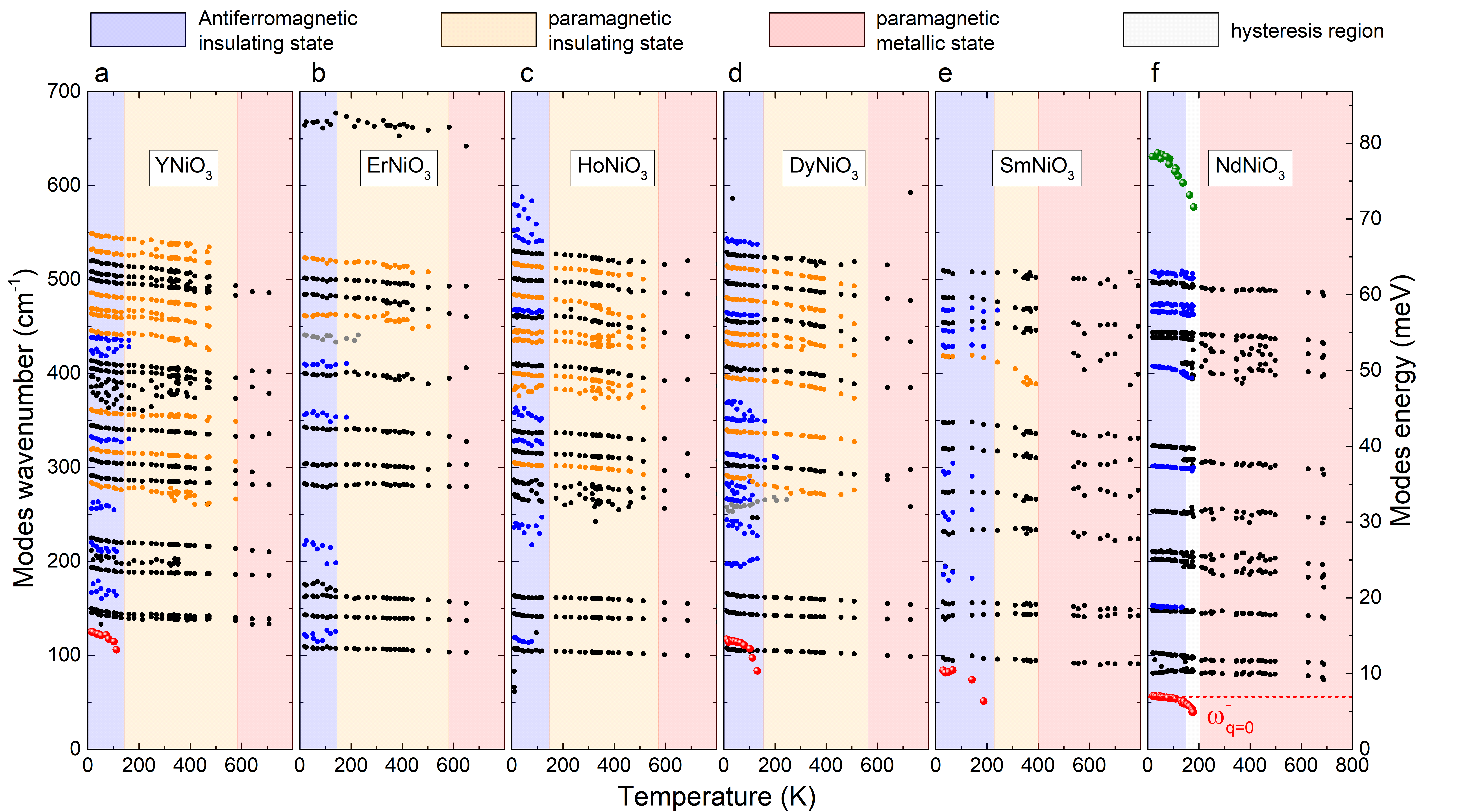}
\caption{\label{fig:2}
Evolution with temperature of the frequency of the Raman modes. 
Black symbols: Modes already present above $T_{MI}$.
Orange symbols: Additional modes below $T_{MI}$. 
Blue symbols: Additional modes below $T_N$. 
Red symbols: Anti-ferromagnetic resonance. 
Grey symbols for ErNiO$_3$ and DyNiO$_3$: Additional modes appearing halfway $T_{MI}$ and $T_N$.
Green symbols (NdNiO$_3$): Bimagnon bound state (tentative, see discussion).}
\end{center}
\end{figure*}

Various different types of magnetic order have been inferred from neutron diffraction, soft x-ray resonant diffraction and resonant inelastic x-ray scattering~\cite{garcia1994,fernandez2001,scagnoli2008,munoz2009,meyers2015,lu2018}, all of them characterized by the lack of an inversion center, but differing in the collinear or non-collinear arrangement of the magnetic moments associated to the two types of (short- and long-bond) nickel sites. 
Theoretically it has been predicted that the combination of breathing distortion and non-centrosymmetric magnetic order can induce a breaking of the inversion symmetry also in the crystal lattice and thus induce an electric dipole, making them type-II multiferroics.~\cite{cheong2007,brink2008,catalan2008,giovannetti2009,hao2013,xin2014,khomskii2017,perez2016,xin2020,spaldin2019}.

\section{Results for the paramagnetic phase}
Small single crystals of RNiO$_3$ with R = Y, Er, Ho, Dy, Sm, Nd were prepared as described in Ref.~\onlinecite{gawryluk2020}. 
A DyNiO$_3$ crystal is shown in Fig.~\ref{fig:1}(a). 
We measured the Raman spectra of RNiO$_3$ crystals with six different compositions in a broad range of temperatures. We monitored the change of the spectroscopic features as the materials pass through three different phases, namely metallic at high temperatures, paramagnetic insulating at intermediate temperatures, and magnetically ordered at the lowest temperatures. 
The experimental procedures are detailed in Appendix A.  
The Raman spectra of DyNiO$_3$ are shown in Fig.~\ref{fig:1}(b) for representative temperatures in the metallic phase (729 K), deep inside the paramagnetic insulator phase (205 K) and deep inside the magnetically ordered phase (12 K). 
In Fig.~\ref{fig:1}(c) the full dataset of DyNiO$_3$ covering the entire temperature range is represented as a colormap of the Raman response  in the frequency-temperature plane (see Appendix A for the other samples).  
For all samples the evolution of the phonon frequencies as a function of temperature is summarized in Fig.~\ref{fig:2}. 
We begin with an observation on the character of the Raman modes. 
In the metallic and insulating phases the crystal point group has, at least down to $T_N$, inversion symmetry. 
Consequently the optical modes are either Raman active (even parity) or dipole active (odd parity).  
The $Pbmn$ (metallic phase at high temperature) and the $P2_1/n$ (insulating phase at low temperature) space groups contain four formula units, and therefore have 60 vibrational degrees of freedom corresponding to 3 acoustic modes, 33 odd parity modes observable in the optical conductivity and 24 even parity modes that are Raman-active. 
Another important feature shared by the $P2_1/n$ and $Pbmn$ phases is that the four Ni atoms are at centers of inversion symmetry. 
Consequently only the motions of oxygens and rare earth atoms contribute to the Raman modes.

The overall trend of the mode frequencies as a function of rare earth element (Fig.~\ref{fig:2}) is a blueshift of all frequencies when moving from the heaviest (Nd) to the lightest (Y) element. 
This trend is most pronounced for the three main modes below 230~cm$^{-1}$, where in particular the case of YNiO$_3$ stands out in that the three main modes have frequencies approximately 35-40~\% higher than for HoNiO$_3$, in agreement with the ratio $(m_{Ho}/m_Y)^{0.5}=1.36$ expected for vibrations with a rare earth character mainly.
The modes above 500~cm$^{-1}$ shift up by approximately 10~\% from Nd to Y, revealing a hardening of the chemical bond, a trend that is expected in view of the reduction of the lattice constant for the smaller rare earth elements.
Cooling through $T_{MI}$ the four NiO$_6$ octahedra within the primitive cell bifurcate into one set with short Ni-O bonds and a second set with long bonds.
While the number of Raman active modes (24) is not affected by the symmetry breaking, we may reasonably expect that intensity and frequency of these modes change at the transition. 
For all RNiO$_3$ compounds of this study the metal-insulator phase transition results in a significant blueshift of all modes (Figs.~\ref{fig:1}, ~\ref{fig:2} and ~\ref{fig:6}).
Above $T_{MI}$ the vibrational features are much broader than below $T_{MI}$. Consequently a number of modes that are close in frequency can be separated from each other below $T_{MI}$ (marked in orange in Fig.~\ref{fig:2}) but not above $T_{MI}$.
A weak mode at 450~cm$^{-1}$ for ErNiO$_3$ and one at 260~cm$^{-1}$ for DyNiO$_3$ broaden progressively as a function of increasing temperature (grey symbols in Fig.~\ref{fig:2}) and become indistinguishable midway $T_{N}$  and $T_{MI}$. The fact that they show up far above $T_{N}$ indicates that these are Raman active modes of the paramagnetic phase.

\begin{figure}[!t]
\begin{center}
\includegraphics[width=1.0\columnwidth]{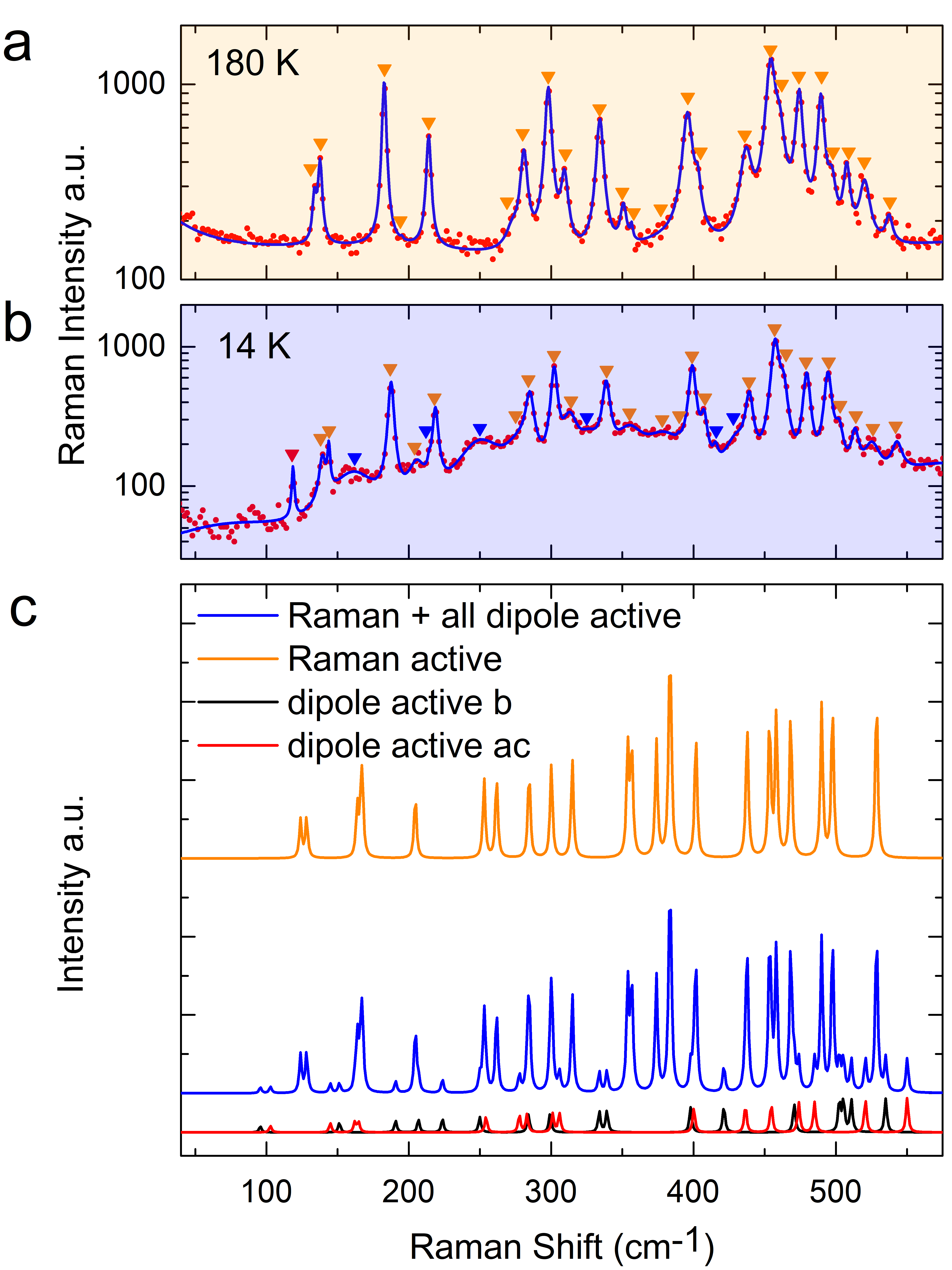}
\caption{\label{fig:3}
Raman spectra of an YNiO$_3$ single crystal (a), in the paramagnetic insulating phase (180~K) and (b), in the antiferromagnetic insulating phase (14~K). 
Orange triangles: Raman modes of the paramagnetic phase. 
Blue triangles: Additional modes in the magnetic phase. Red: anti-ferromagnetic resonance.
(c), simulated Raman spectrum of YNiO$_3$ in the paramagnetic insulating phase (orange) and the insulating antiferromagnetic phase (blue) where it is assumed that all dipole active phonons ($b$-axis: black, $ac$-plane: red) have become Raman active as a result of inversion symmetry breaking. 
The calculation of the phonon frequencies is described in Appendix B.
The modes have a Lorentz line-shape with a full-width at half-maximum of 4~cm$^{-1}$. The amplitudes $f_j$ of the Raman modes (see Eq.~\ref{eqnSum}) were all given the same value, and the amplitudes of all polar modes 1/5th of those of the Raman modes. }
\end{center}
\end{figure}
\begin{figure*}[!ht]
\begin{center}
\includegraphics[width=1.3\columnwidth]{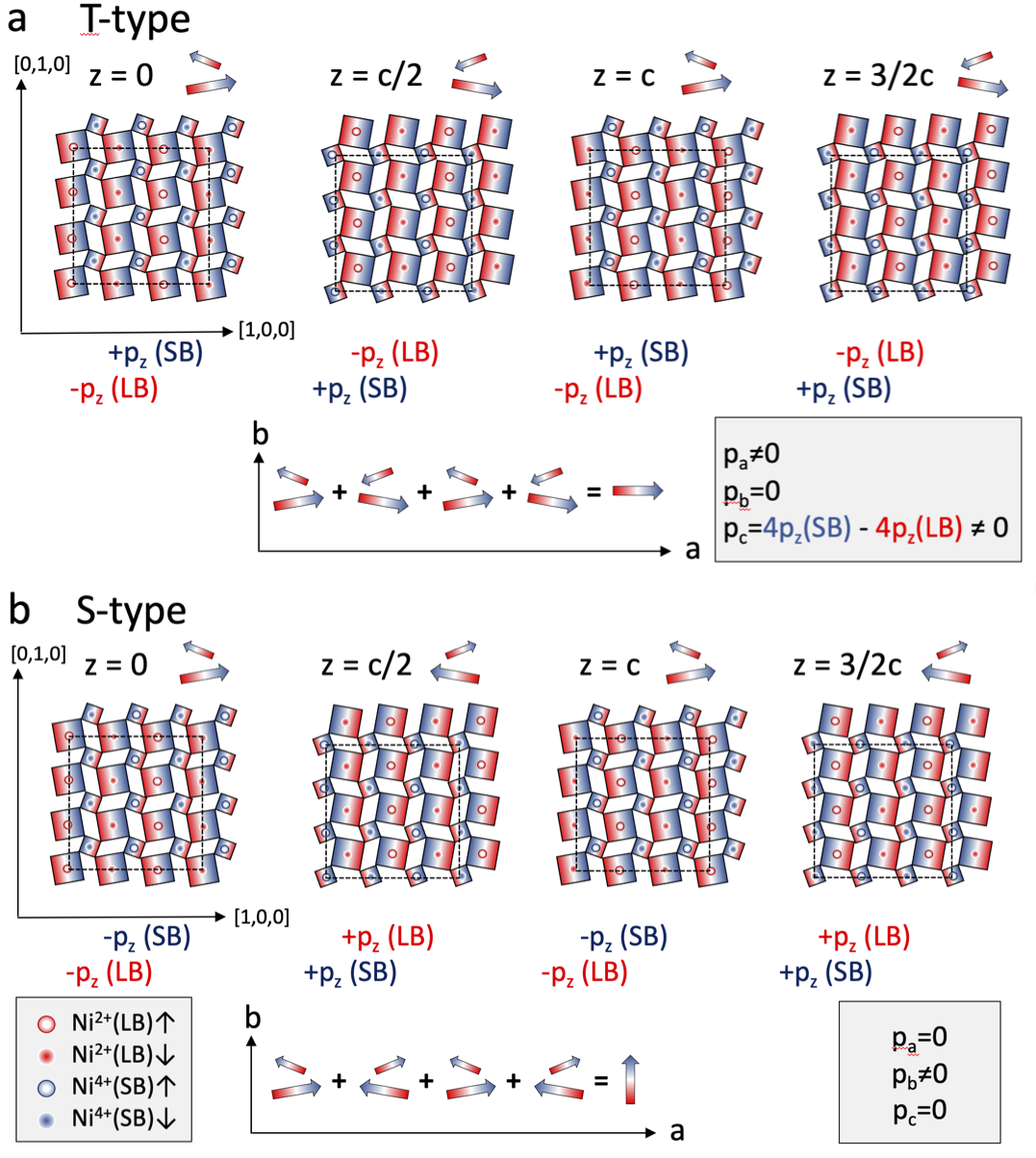}
\caption{\label{fig:4}
Multiferroicity of {\bf a} $T$-type and {\bf b} $S$-type antiferromagnetic order~\cite{giovannetti2009}.  
Coordinate axes are labeled according to $P2_1/n$ setting.
Dotted squares indicate the alignment of the planes stacked along  $[0,0,1]$. 
Labels for up/down spin and long/short-bond Ni are given in the top right. }
\end{center}
\end{figure*}
\section{Results for the antiferromagnetic phase}
Cooling the samples through the magnetic phase transition has a surprisingly large impact on the Raman spectra, in particular showing a large number of additional modes (see Fig.~\ref{fig:2}) which we interpret as phonons, with a few exceptions that we will discuss first: For all compounds except ErNiO$_3$ and HoNiO$_3$ we observe an additional mode at the low frequency end of the spectrum, showing a strong redshift upon raising the temperature to $T_N$. 
For NdNiO$_3$ the frequency is 58~cm$^{-1}$ at the lowest temperatures.
Since this is in the range of the zone-center magnon obtained for the same material from extrapolating resonant inelastic x-ray scattering (RIXS) data (13 meV)~\cite{lu2018} we interpret this mode as  the anti-ferromagnetic resonance.
NdNiO$_3$ is the only compound showing a peak near 640 cm$^{-1}$ with a strong redshift upon raising the temperature to $T_N$. This is probably a bi-magnon: Taking the experimental value of the energy of the zone-boundary magnon from RIXS~\cite{lu2018} we arrive at $2\times 55$~meV giving 887~cm$^{-1}$ as the upper bound. The lower frequency in the Raman spectrum can be qualitatively understood as a consequence of magnon-magnon coupling.  

As is illustrated in Fig.~\ref{fig:1}(b) for the case of DyNiO$_3$, we observe below $T_N$ at least 5 relatively strong additional modes at 120, 270, 310, 350, 460~cm$^{-1}$ and 2 weaker ones at 200 and 540~cm$^{-1}$. 
The total number of modes in the low temperature Raman spectrum is 22 out of 24 that are Raman active, {\em i.e.} the number of Raman active modes observed in DyNiO$_3$ does not indicate as such that the space group in the magnetic phase should be different from $P2_1/n$. However, it is remarkable that certain modes appear in pairs. 
For DyNiO$_3$ the modes near 305, 340 and 450~cm$^{-1}$ split into doublets below $T_N$. 
The splitting of the 340~cm$^{-1}$ mode is also clearly seen for HoNiO$_3$ and ErNiO$_3$ and for some of the other modes in specific samples. 
NdNiO$_3$ is special among the samples studied in that we see for all phonons an abrupt narrowing and an intensity change at $T_N$ (=$T_{MI}$), and  the 305~cm$^{-1}$ mode splits into a strong peak at 300 ~cm$^{-1}$ and a weak one at 315~cm$^{-1}$, confirming the observations in Refs.~\onlinecite{girardot2008,zaghrioui2001}.
For DyNiO$_3$ a well defined mode appears near 280~cm$^{-1}$ surrounded by two shoulders. 
The same mode is observable at low temperature for HoNiO$_3$ with the difference that this narrow peak is still present as a shoulder in the paramagnetic phase. 
We observe also the appearance of a broad mode at 250~cm$^{-1}$, visible for YNiO$_3$, HoNiO$_3$ and DyNiO$_3$ (Fig.~\ref{fig:2}).  
A second well defined characteristic mode appears at around 350~cm$^{-1}$ for HoNiO$_3$, DyNiO$_3$ and ErNiO$_3$. 
Other new modes emerge at 550 and 570~cm$^{-1}$ for HoNiO$_3$, at 200 and 550~cm$^{-1}$ for DyNiO$_3$ and at 440~cm$^{-1}$ for ErNiO$_3$. 

For the interpretation of the additional modes below $T_N$ the case of YNiO$_3$ is of particular importance. 
Taking a close look at the Raman spectra in the paramagnetic and antiferromagnetic insulating phase shown in Fig.~\ref{fig:3}(a,b), we notice that (i) there are 24 Raman modes above $T_N$ and (ii) this number increases to 31 below $T_N$. 
One of the extra modes (the one at 120 cm$^{-1}$) is the antiferromagnetic resonance. 
Since $Y$ has no $4f$ electrons, we can exclude beyond any doubt contributions of $4f$ crystal field excitations to the Raman spectra. 
The only remaining interpretation of the additional modes is therefore that these are phonons. 
Since for $P2_1/n$ the number of Raman active phonons is limited to 24, observing 30 Raman active phonons is a clear demonstration that the space group in the antiferromagnetic phase must be different from $P2_1/n$. 
%Some of the peaks, {\em e.g.} the additional mode around 270 cm$^{-1}$ in DyNiO$_3$ in Fig.~\ref{fig:1}(b), are already present above $T_N$. This is possibly the consequence of short range antiferromagnetic order that is still present above $T_N$. This interpretation is consistent the disappearance of intensity within a range of 20~K as the temperature is increased above $T_N$.

\section{Discussion}
\begin{figure}[!t]
\begin{center}
\includegraphics[width=0.7\columnwidth]{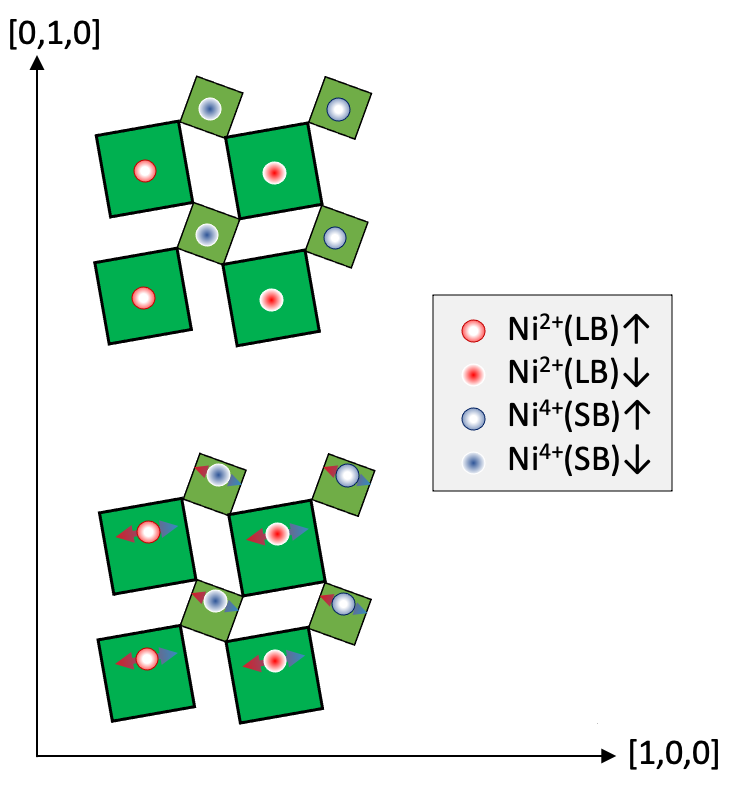}
\caption{\label{fig:5}
Scheme of the spontaneous inversion symmetry breaking. 
Labels for up/down spin and long/short-bond Ni are indicated at the right hand side. 
To avoid clutter oxygen atoms (at the corners of the squares) and rare earth atoms (in the open spaces) are not indicated. 
%The atomic displacements of the $\Gamma^{4-}$ and $R^{4-}$ modes are illustrated for the Ni ions. The 
Top: The Ni ions occupy inversion symmetric positions. 
Bottom: The electric fields induced by the T-type or S-type antiferromagnetic order (see Fig.~\ref{fig:4}) push the Ni ions to low symmetry positions, which has two main effects: (i) It contributes a net electric dipole moment. (ii) Since these positions lack inversion symmetry, modes of any polarization are simultaneously Raman active and dipole active. Vibrations around the rest-position are symbolized by double-headed arrows.}
\end{center}
\end{figure}

It is fully established that the magnetic order of these compounds is described by the propagation vector (1/2,0,1/2) (or (1/4,1/4,1/4) when referred to the underlying pseudocubic lattice~\cite{footnote1}), but the orientation of the magnetic moments is less clear~\cite{garcia1994,fernandez2001,scagnoli2008,munoz2009,meyers2015,lu2018}. 
Giovannetti {\em et al.}~\cite{giovannetti2009} discuss three types of magnetic order labeled $T$, $S$ and $N$. For $T$-type order, the magnetic moments are arranged in planes of parallel spins alternating as $\uparrow\uparrow\downarrow\downarrow$ along the $[1,1,1]$ direction of the underlying pseudocubic structure, which they predict to be polar with electric dipole components along $a$ and $c$. 
For $S$-type, they predict a net electric dipole along $b$. The predicted multiferroicity for $T$ and $S$ originates in the phenomenon that theoretically regions between parallel (opposite) magnetic moments become electron poor (rich) due to exchange-correlation.  
As illustrated in Fig.~\ref{fig:4} this would result in the appearance of a net electric moment, whose direction can be along the $b$-axis or perpendicular to it depending on the type of magnetic order.  
The non-collinear magnetic structure $N$ owes its multiferroicity to spin-orbit coupling, and has an electric dipolar moment much weaker than the collinear $T$ and $S$ arrangements~\cite{giovannetti2009}. 
Perez-Mato {\em et al.} predicted the existence of magnetism-induced ferroelectricity from group theory arguments based in the symmetry of the paramagnetic space group and the magnetic propagation vector~\cite{perez2016}. According to these arguments a symmetry decrease from centrosymmetric $P2_1/n$ to  polar $P2_1$ and the appearance of spontaneous polarization along the $b$ crystal axis are expected to occur below $T_N$ for the collinear and non-collinear magnetic structures reported respectively for PrNiO$_3$~\cite{garcia1994} and HoNiO$_3$~\cite{fernandez2001}. 
In any of these scenarios the electric fields induced by the electronic charge displacements will act as a force on the ions, deforming the lattice and -most significantly- pushing the Ni-ions away from centrosymmetric positions. 
%The possibility of a spontaneous polarization along the $b$ axis is supported by the recent observation of lattice anomalies at the setup of the antiferromagnetic order, which are particularly pronounced for the $b$ lattice parameter~\cite{gawryluk2020}, and the additional distortion modes (also called {\textquotedblleft}frozen phonons{\textquotedblright})  belonging to the irreducible representations $\Gamma_4^{-}$ and $R_4^{-}$ have been identified as the most likely atomic displacements at the origin of the observed lattice anomalies. 
%The two only modes belonging to these irreducible representations that involve displacements of the Ni atoms along $b$ are schematically shown in Fig.~\ref{fig:5} together with their impact on the optical selection rules.
As illustrated in Fig.~\ref{fig:5} this relaxes the optical selection rules. In particular it causes dipole active vibrations to become simultaneously Raman active and vise versa.

\section{Key implications of the experiments}
Polarized neutron diffraction experiments have proven the magnetic origin of the aforementioned $(1/2,0,1/2)$ superlattice~\cite{garcia1994,rodriguez1998,fernandez2001}. Consequently the magnetic order modulates the force constants with the same lattice periodicity as the one of the paramagnetic state, thus excluding the possibility that the additional modes are folded zone-boundary phonons. 
If on the other hand a breaking of inversion symmetry were to occur, this would have a major impact on the Raman spectra. 
The implication for the vibrational modes is that all dipole active modes become simultaneously Raman active and vice versa.
For non-polarized Raman as used for the spectra in Fig.~\ref{fig:3} at most 33 polar modes will then contribute to the Raman spectra below $T_N$. 
The {\em ab initio} calculations (see Appendix B) of the vibrational spectrum presented in Fig.~\ref{fig:3}(c) demonstrate that the features in the experimental data of Fig.~\ref{fig:3}(b) are indeed consistent with polar modes that become Raman active below $T_N$. 
Some of the additional modes below $T_N$ show up as shoulders of peaks already present in the paramagnetic insulating phase.
The fact that some of these modes appear as pairs and others not, is then a simple consequence of the fact that some of the Raman and polar modes correspond to similar vibrational patterns, the main difference being the in-phase and out-of-phase motion of subsets of the atoms (in the present case four formula units) inside the primitive cell. 
Furthermore we observe a total of 30 vibrational modes at low temperature for YNiO$_3$ which is 6 more than allowed by the factor group analysis. The implication from these results is that the antiferromagnetic phase is polar. 

Moving from Nd to Y, the number of additional modes that we can detect below $T_N$ increases. 
That the multiferroic effect is strongest for the smaller rare earth ions has in fact been anticipated by Giovannetti {\em et al.}~\cite{giovannetti2009}. 
The ingredients responsible for the multi-ferroic effects are magnetic order, breathing distortion and rotation of the oxygen octahedra (Fig.~\ref{fig:4}). 
This combination of factors is uncommon in the multiferroic family tree~\cite{spaldin2019} and characteristic for the rare earth nickelates. 
Of the samples studied here YNiO$_3$ has the smallest tolerance factor (Table~\ref{table:crystals}) and consequently the largest octahedral rotation, highest $T_{MI}$ and the strongest breathing distortion. 
While the moderate decrease of $T_N$ from NdNiO$_3$ to YNiO$_3$ suggests a weakening of the magnetic order parameter, this is more than compensated by the increasing octahedral rotation and breathing distortion, together resulting in the observed positive trend of the multiferroic coupling from large to small tolerance factor. 

\begin{table}[!t]
\begin{center}
\begin{tabular}{| c | c | c | c | c | c |c | c | c |}
\hline
Compound  & $T_N$  & $T_{MI}$ & $r_{RE}$ & $t$ \\
 & K       & K            &    & \\
\hline
YNiO$_3$ &143 &584 &1.019 & 0.862 \\
ErNiO$_3$ &145 &582 &1.004 & 0.864 \\
HoNiO$_3$ &147 &573 &1.015 & 0.866 \\
DyNiO$_3$ &155 &564 &1.027 & 0.868 \\
SmNiO$_3$ &229 &401 &1.079 & 0.894 \\
NdNiO$_3$ &203 &202 &1.109 & 0.915 \\
\hline
\end{tabular}
\caption{\label{table:crystals} 
N\' eel temperature~\cite{catalan2008}, insulator-metal transition temperature~\cite{catalan2008}, rare earth radius (8-fold coordination)~\cite{shannon1976}, and tolerance factor~\cite{catalan2008} of the RNiO$_3$ compounds that are subject of the present study.}
\end{center}
\end{table}

We see that the Raman spectra provide experimental support for the predicted existence of magnetism-driven ferroelectricity, demonstrating the clear symmetry breaking below $T_N$ and confirms previous indirect experimental evidence supporting this idea. For PrNiO$_3$, some of us analyzed the temperature dependence of the crystal structure in terms of symmetry adapted modes~\cite{gawryluk2019}. 
The unconventional temperature dependence of the lattice parameters below $T_{MI}= T_N$, and the observation of a nearly perfect linear correlation between the breathing mode amplitude and the staggered magnetization, lead to the conjecture of a hidden order in the insulating phase, in particular the predicted polar distortion induced by the noncentrosymmetric magnetic order~\cite{giovannetti2009,perez2016}. 
The clear lattice anomalies recently observed at T$_N$ for EuNiO$_3$ \cite{serrano2021} confirm this intuition, giving additional support to the Raman findings.
In summary it has been anticipated theoretically that magnetic order in RNiO$_3$ breaks inversion symmetry, corresponding to a multiferroic state in which all dipole active phonons are simultaneously Raman active. 
Our observation of additional Raman active modes below $T_N$ provides experimental for this prediction.
Full characterization of the phase below $T_N$ requires additional experimental investigations such as second harmonic generation, measurements of the electric dipole field and crystallographic experiments.

The datasets generated and analyzed during the current study are available in Ref.~\onlinecite{yareta2021} and will be preserved for 10 years. 

\section{Acknowledgements}
 This project was supported by the Swiss National Science Foundation through projects 200020-179157, 200021-163103, 200020-185061, NCCR MARVEL (Grant No. 51NF40-182892), and R\'equip Grant No. 206021-139082.
\appendix
\begin{figure}[!t]
\begin{center}
\includegraphics[width=0.8\columnwidth]{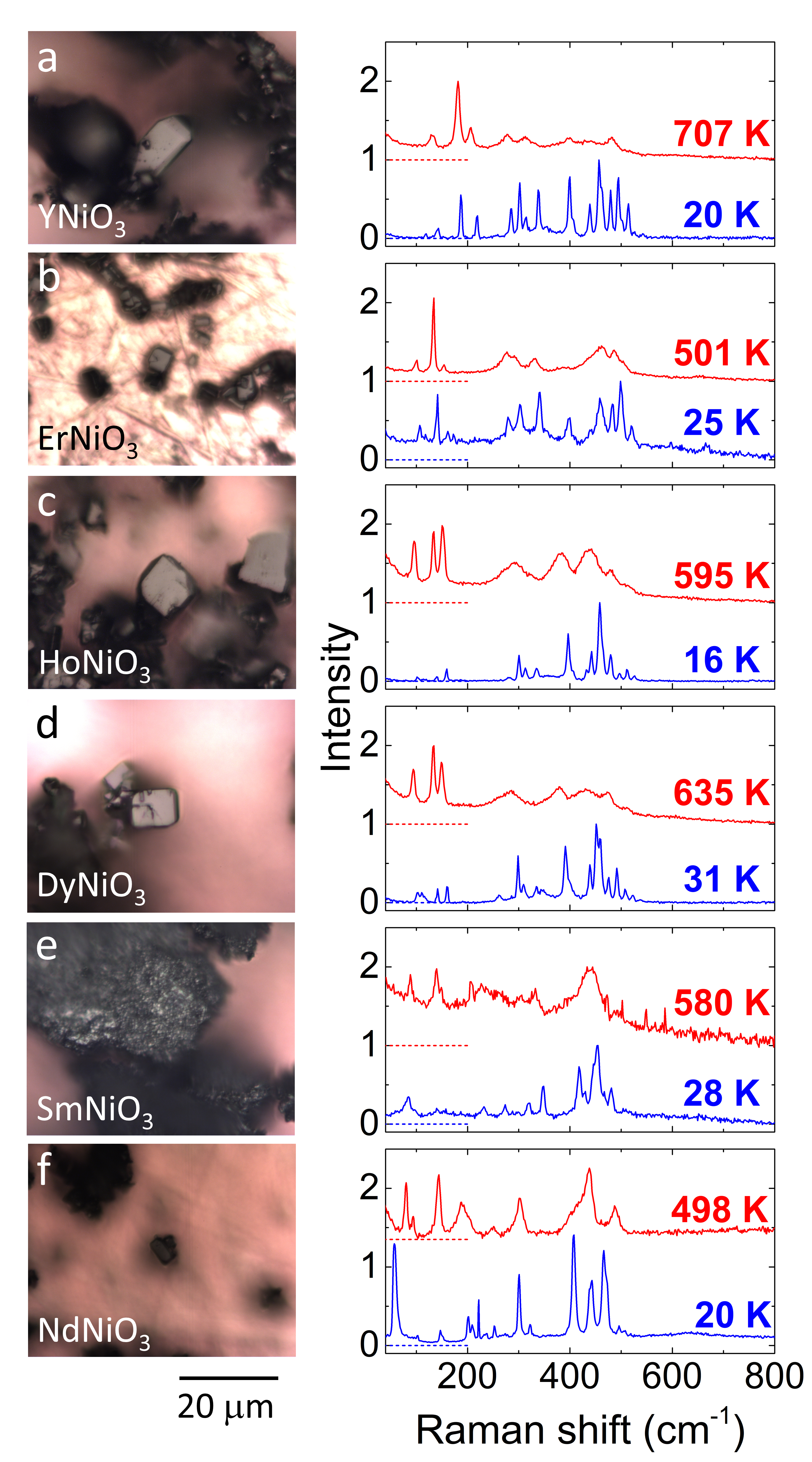}
\caption{\label{fig:6}
Images of {\bf a}, YNiO$_3$, {\bf b}, ErNiO$_3$, {\bf c}, HoNiO$_3$, {\bf d}, DyNiO$_3$, {\bf e}, SmNiO$_3$ and {\bf f}, NdNiO$_3$ single crystals inside the cryostat and the associated raw Raman spectra in the high temperature metallic state (red curve) and in the low temperature antiferromagnetic state (blue curve). 
Temperature of each spectrum is indicated on the curve. 
Scale bar is identical for all samples. }
\end{center}
\end{figure}
\begin{figure*}[!t]
\begin{center}
\includegraphics[width=1.5\columnwidth]{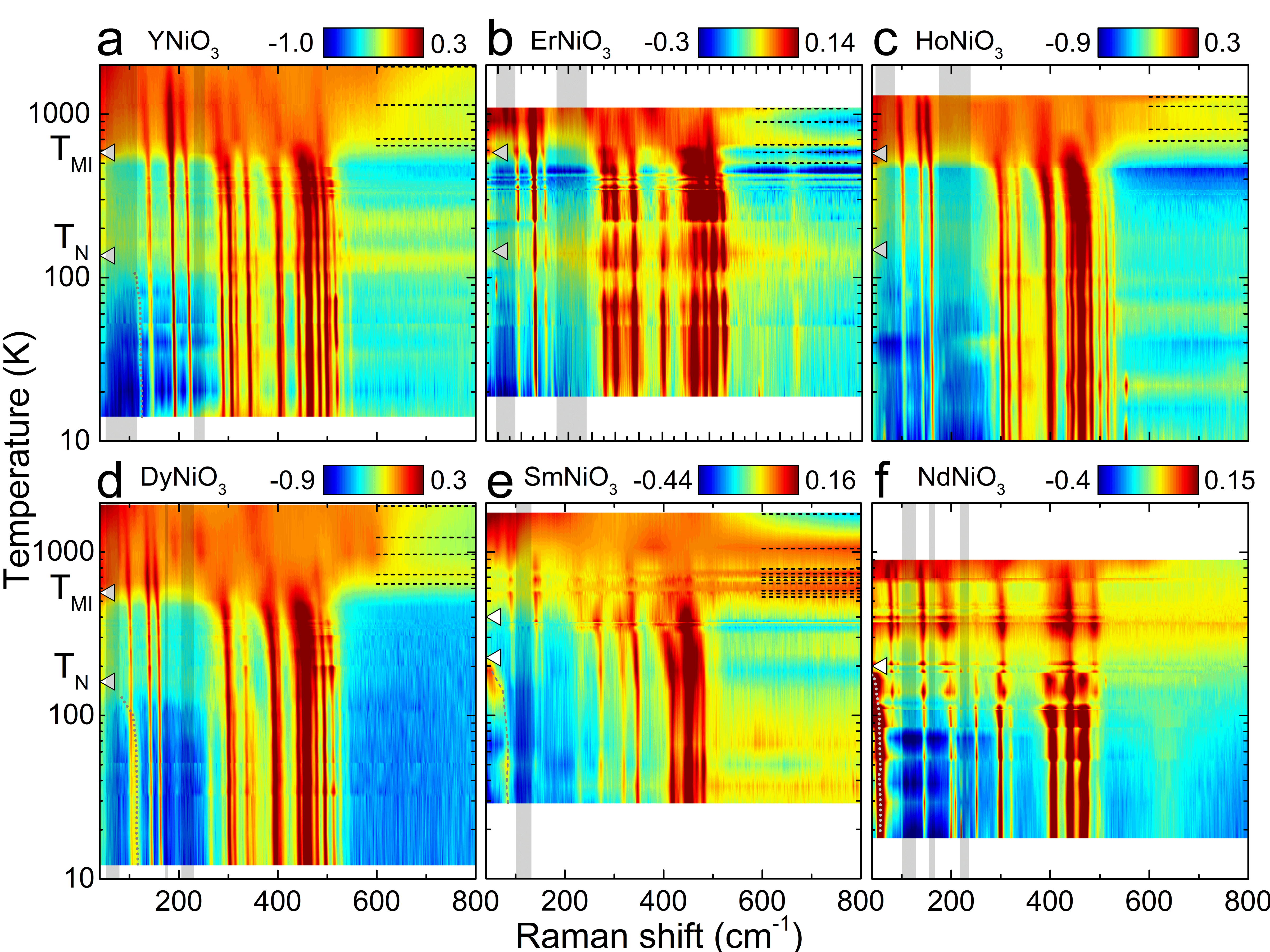}
\caption{\label{fig:7}
Colormap of Raman intensity for {\bf a}, YNiO$_3$, {\bf b}, ErNiO$_3$, {\bf c}, HoNiO$_3$, {\bf d}, DyNiO$_3$, {\bf e}, SmNiO$_3$ and {\bf f}, NdNiO$_3$ samples. 
Horizontal axis is the Raman shift in~cm$^{-1}$, vertical axis is the temperature in K on a log scale, color defined the intensity of the signal in a logarithmic scale. 
White triangles indicate the temperatures of metal to insulator (high temperature) and magnetic ordering (low temperature). 
When present, the soft-mode frequency from the spectral fit is marked with a dashed gray line. 
Gray regions are averaging windows used for the temperature dependence of the Raman electronic background. 
Dotted horizontal lines indicate the temperatures of the spectra above 700 K used for the color plot.The temperature mesh below 700 K is too dense to indicate this way. }
\end{center}
\end{figure*}
\section{Experiments}
The micro-crystals~\cite{gawryluk2020} were glued on the copper plate of a He flow cryostat (Konti Micro from CryoVac GMBH) using high vacuum grease. 
Raman spectra were recorded using a Horiba LabRAM HR Evolution spectrometer with an excitation wavelength of 532~$nm$ and the resolution of $2$~cm$^{-1}$ over the full range. 
The laser light was focused on a $2~\mu m$ spot using a window corrected $63\times$ objective. 
Size, shape and orientation of the incident polarisation (parallel to the horizontal axis) varied from one sample to another. 
Images of some of the single crystals mounted in the cryostat are displayed in Fig.~\ref{fig:6}, together with the corresponding Raman spectra in the high temperature metallic phase (red curves) and in the low temperature antiferromagnetic phase (blue curves). 
The small size of the crystals prohibited reproducible alignment and concomitant polarization analysis of the Raman spectra.
 
Spectra were measured in a broad temperature range from 10~K to 1000~K for most samples, and for NdNiO$_3$ up to 1911~K.  
The variable power of the excitation laser was used for heating the micro-crystals. 
The temperature of the sample is determined by the equilibrium of heat exchange between the crystal and the sample holder and heating by different power of the laser. 
Due to differences in, amongst other,  absorption coefficient, size, shape and the contact between grease layer, crystal and copper, the thermal coupling of the crystals to the cold plate varies from one specimen to another. 
For these reasons the same incident laser power heats up the different crystals to different temperatures. 
The sample temperature was calculated from the Stokes/anti-Stokes ratio as detailed below. 
As the laser power had to be varied in order to tune the sample temperature, it was necessary to normalize the spectra presented in Fig.~\ref{fig:1} of the main text and Fig.~\ref{fig:7}. 
For the normalization, firstly the average intensity of the Stokes response was extracted between 250 and 550~cm$^{-1}$ and each spectrum was normalized to that average intensity. 
In a second step, we corrected for the drift between subsequent temperatures. 
To that end, for each temperature we extracted the average Stokes response in a frequency range without Raman active modes and as a function of temperature fitted these averages using a polynomial. 
Then, for each temperature, the difference between the extracted average and the fitted value was subtracted from the entire Raman spectrum, effectively removing background jitter as a function of temperature. 

\begin{figure}[t!]
\begin{center}
\includegraphics[width=1\columnwidth]{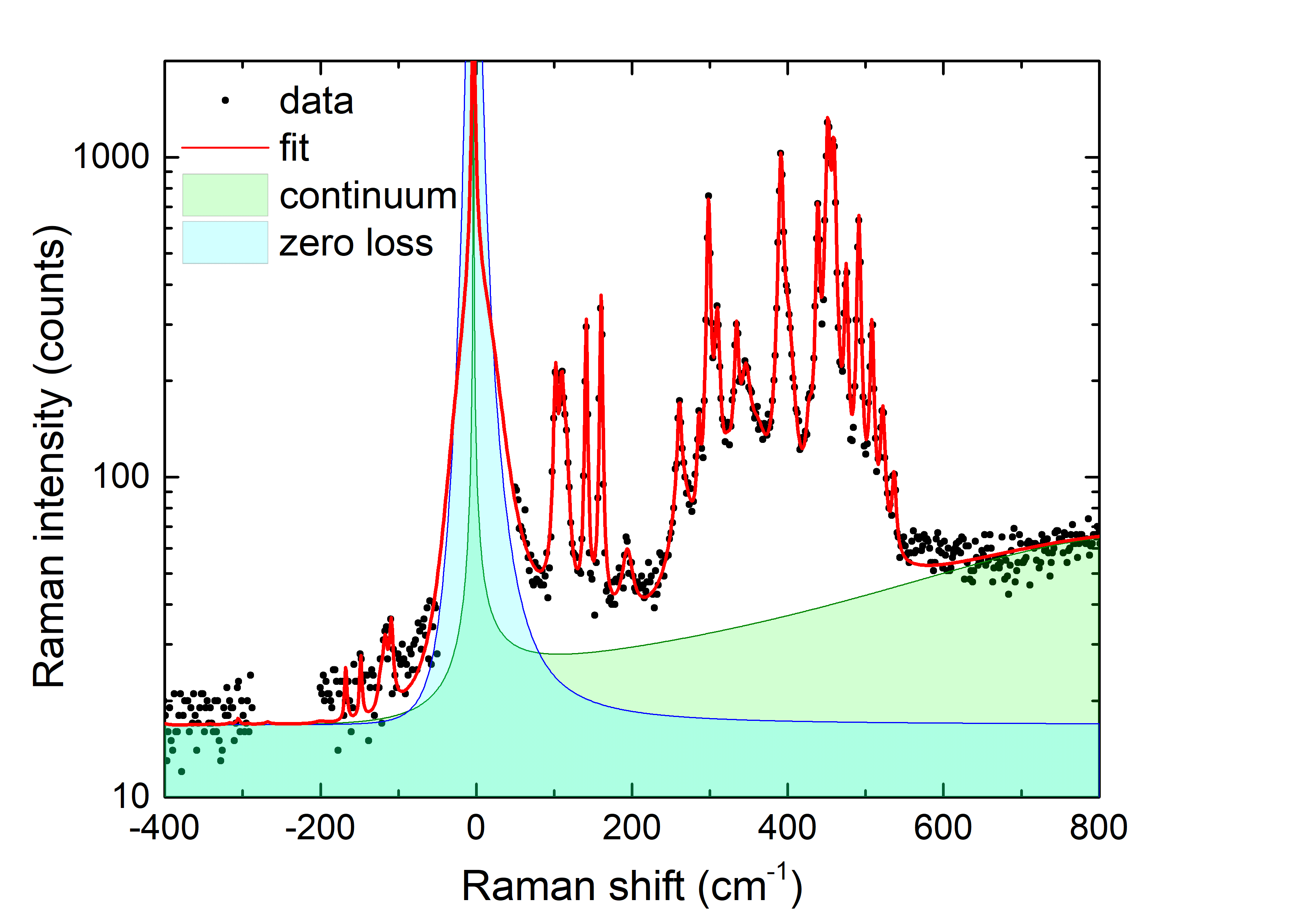}
\caption{\label{fig:8}
Raman spectrum of DyNiO$_3$ shown together with a multi-oscillator fit obtained by adjusting the temperature, the oscillator parameters of Eq.~\ref{eqnSum}, a background and the zero of frequency. }
\end{center}
\end{figure}

The Raman spectrum at finite temperatures is given by
\begin{equation}\label{eqnBWF}
I\left(\omega,T\right) = \frac{\mbox{Im}\chi(\omega)}{1-e^{-\hbar\omega / k_B T}}
\end{equation}
 where $\omega=\omega_{in}-\omega_{out}$ is the Raman frequency shift, $\omega_{in}$ ($\omega_{out}$) is the frequency of the incoming (scattered) photon, $T$ is the temperature, $k_B$ the Boltzmann constant and $\chi(\omega)$ the Raman susceptibility. The imaginary (real) part of $\chi(\omega)$ is an odd (even) function of $\omega$. Since the denominator of Eq.~\ref{eqnBWF} changes sign at the origin of $\omega$, the function $I\left(\omega,T\right)$ is positive for positive and negative $\omega$. 
Positive (negative) frequencies represent the Stokes (anti-Stokes) side of the spectrum. Since the Stokes/anti-Stokes ratio 
\begin{equation}\label{SaS}
\frac{I\left(\omega,T\right)}{I\left(-\omega,T\right) }=  e^{\hbar\omega / k_B T}
\end{equation} 
is a universal function of $\omega/T$, the temperature can in principle be readily obtained from this ratio. In practice one first needs to correct the spectra for the zero-loss peak and for a dark current background. In addition it is necessary to correct for spectrometer drift of the zero of the frequency axis. Since Eq.~\ref{SaS} can only be applied after these corrections have been made, we use a different approach, which consists of fitting~\cite{kuzmenko2005} the spectra for positive and negative Raman shifts to a sum of oscillators $\chi_j(\omega)$ for which we use 
\begin{equation}\label{eqnSum}
\chi( \omega )  =  \sum_{j} 
\frac{\omega^2 f_j }{\omega_j^2-\omega(\omega+i\gamma_j)}
\end{equation} 
 where $f_j$ parametrizes the amplitude of the $j$'th Raman mode, $\omega_j$ it's peak frequency, and $\gamma_j$ it's half width at half maximum. For all samples the spectra at different temperatures could be satisfactory modeled with a Lorentzian line shape for the oscillators. 
The temperature of the sample is used as fitting parameter together with the parameters $\omega_j$, $\gamma_j$ and $f_j$ of the oscillators. 
Additional adjustable parameters were the zero of the frequency axis, a frequency independent background, and a symmetric profile for the zero-frequency peak. 
We excluded the frequency range $-50$~cm$^{-1}<\omega<50$~cm$^{-1}$ from the fitting procedure for all crystals except NdNiO$_3$ for which the excluded range was $-20$~cm$^{-1}<\omega<20$~cm$^{-1}$. 
Crucially, the Stokes and anti-Stokes side of the spectrum were acquired simultaneously and the spectrum for $\omega<0$ and $\omega>0$ was fitted with a single set of parameters. 
In Fig.~\ref{fig:8} the example of DyNiO$_3$ is shown. The sample holder was kept at 20~K. The relative intensity of the Stokes and anti-Stokes peaks reveals that the temperature at the location of the laser beam is 64~K.

\section{Phonon Calculations}
The phonon frequencies presented in Fig.~\ref{fig:3}{\bf c} of the main text were calculated using Density-Functional. Perturbation Theory as implemented in Quantum Espresso~\cite{giannozzi2009}. 
The monoclinic unit cell was relaxed and an all electrons self consistant calculation was used with a generalized gradient approximation within the Perdew–Burke–Ernzerhof functional~\cite{perdew1996}. 
Cut-off energies for kinetic energy and charge density were 100~Ry and 600~Ry respectively. The $k$-points corresponding to the electronic wave functions were integrated on a $7\times 6\times 5$ Monkhorst–Pack grid yielding $72$~$k$-points in the irreducible wedge of the Brillouin zone. Dynamical matrices were then computed using a self consistent convergence criterion of $10^{-12}$~Ry.
The structure was relaxed to self consistent convergency, with the effect that lattice parameters do not exactly match the experimental values.

\section{Magnetic space group and multiferroicity}
The magnetic propagation vector of the RNi$O_3$ expressed on the reciprocal lattice basis of the $P2_1/n$ space group is $(1/2,0,1/2)$. $P2_1/c$ describes the same space group and has the same $\vec{a}$ and $\vec{b}$ primitive lattice vectors, but $\vec{c}_{c}$ of $P2_1/c$ equals $\vec{a}+\vec{c}_{n}$ of $P2_1/n$. 
Correspondingly the  magnetic propagation vector expressed on the reciprocal lattice basis of $P2_1/c$ is $(1/2,0,0)$, {\em i.e.} the $P2_1/c$ setting reveals that this magnetic order  doubles the volume of the non-magnetic primitive cell. 
In Hermann-Mauguin notation the symbol $2_1$ refers to a 2-fold screw axis parallel to the unique ($b$) axis of the monoclinic structure and the symbol $/c$ to a glide plane, more specifically a mirror operation in a plane perpendicular to $b$  followed by a translation along $\vec{c}_{c}/2$~\cite{bradley2010,litvin2013}. 
If the glide plane symmetry is broken, the corresponding magnetic space group (MSG) $P_{2a}2_1$ supports a net electric dipole along $\vec{b}$~\cite{litvin2013}. 
Likewise, for MSG $P_{2a}c$ the screw axis symmetry is broken, resulting in  a net electric dipole with components along $\vec{a}$ and $\vec{c}_{c}$~\cite{litvin2013}. 
$S$-type and $T$-type order belong to MSG $P_{2a}2_1$ and $P_{2a}c$ respectively.  
For the microscopic description of multiferroicity in $T$- and $S$-type order we follow Giovannetti {\em et al.}~\cite{giovannetti2009} and use $P2_1/n$ setting; correspondingly from here on $c$ refers to $c_{n}$.
Long-bond (LB) and short-bond (SB) Ni sites alternate along $\vec{a}+\vec{b}$, $\vec{a}-\vec{b}$ and $\vec{c}$, with an octahedral counter-rotation in the $ab$-plane of the two types of sites. Exchange-correlation causes electron-depletion (accumulation) between Ni-neighbors of parallel (opposite) spin. 
In the case of T-type order (Fig.~\ref{fig:4}{\bf a}) the net dipole of the planes together is zero along $\vec{b}$ and finite along $\vec{a}$ and  $\vec{c}$.
In the case of S-type order (Fig.~\ref{fig:4}{\bf b}) the net dipole is zero along $\vec{a}$ and $\vec{c}$, and finite along $\vec{b}$. 
%
%
%
%\bibliographystyle{apsrev4-1}
%\bibliography{ArdizzoneReferences}

\begin{thebibliography}{56}
\bibitem [{foo()}]{footnote1}%
  \BibitemOpen
  \href@noop {} {\bibinfo  {journal} {The propogation vector in $P2_1/n$
  setting is $(1/2,0,1/2)$. In $P2_1/c$ setting it is $(1/2,0,0)$,
  corresponding to doubling of the paramagnetic primitive cell along $a$}\
  }\BibitemShut {NoStop}%
\end{thebibliography}
%
%
\end{document}